**Tunable Rashba spin-orbit coupling and its interplay with multiorbital effect and magnetic ordering at oxide interfaces**

Weilong Kong[1, 2, †], Tong Yang[2, †], Jun Zhou[2], Yong Zheng Luo[1], Jingsheng Chen[3], Lei Shen[1], Yong Jiang[4], Yuan Ping Feng[2, 5, *], and Ming Yang[6, *]

[1] Department of Mechanical Engineering, National University of Singapore, Singapore 117575, Singapore

[2] Department of Physics, National University of Singapore, Singapore 117551, Singapore

[3] Department of Materials Science and Engineering, National University of Singapore, Singapore 117575, Singapore

[4] School of Electronics and Information Engineering, Tiangong University, Tianjin 300387, China

[5] Centre for Advanced Two-Dimensional Materials, National University of Singapore, Singapore 117546, Singapore

[6] Department of Applied Physics, The Hong Kong Polytechnic University, Hung Hom, Kowloon, Hong Kong, China

† These authors contribute equally to this work.

∗ Correspondence should be addressed to M.Y. (mingyang@polyu.edu.hk) or Y.P.F. (phyfyp@nus.edu.sg)

**Abstract**

The complex oxide heterostructures such as LaAlO$_3$/SrTiO$_3$ (LAO/STO) interface are paradigmatic platforms to explore emerging multi-degrees of freedom coupling and the associated exotic phenomena. In this study, we reveal the effects of multiorbital and magnetic ordering on Rashba spin-orbit coupling (SOC) at the LAO/STO (001) interface. Based on first-principles calculations, we show that the Rashba spin splitting near the conduction band edge can be tuned substantially by the interfacial insulator-metal transition due to the multiorbital effect of the lowest $t_{2g}$ bands. We further unravel a competition between Rashba SOC and intrinsic magnetism, in which the Rashba SOC induced spin polarization is suppressed by the interfacial magnetic ordering. These results deepen our understanding of intricate electronic and magnetic reconstruction at the perovskite oxide interfaces and shed light on engineering of oxide heterostructures for all-oxides based spintronic devices.

**Introduction**

The essential feature of SOC is that electrons in crystals experience an effective magnetic field in their frame of motion which couples the spin and orbital motion of electrons [1]. Rashba type of SOC emerges at crystal surfaces or interfaces, where the inversion symmetry is broken. It induces spin-momentum locking ($k$-dependent spin polarization) and lifts the spin degeneracy of electronic energy bands (Rashba spin splitting) [2–4]. These properties of the Rashba SOC are appealing in spintronic devices as they facilitate manipulation of the electron spins without the need of an external magnetic field, and even enhance the interconversion between charge and spin currents through Edelstein effect and inverse Edelstein effect [5].

Oxide interfaces are promising candidates for spin-orbit-based control of spintronic devices as they have moderate Rashba SOC and long spin lifetime [6]. A well-known example of oxide interfaces is the LAO/STO (001) interface, which exhibits LAO-thickness-dependent emergent properties that are absent in its bulk counterparts. These include insulator-metal transition [7–9], high-mobility electron gas [10,11], superconductivity [12,13], and interfacial magnetic phase [14–16]. Particularly, gate-tunable Rashba SOC was observed at the LAO/STO interface [17,18], and has been used to demonstrate the interconversion between spin and charge currents [19–22]. Further studies found that the efficiency of the spin to charge currents conversion was dependent on Rashba SOC strength and spin lifetime [6,19,23,24]. Due to longer spin lifetime, the conversion efficiency of the LAO/STO interface is even higher than that of systems with heavier elements like Bi/Ag interfaces or topological insulators such as $Bi_2Se_3$ [6,23,25]. Additionally, the Rashba SOC at the LAO/STO interface has led to many exotic properties, including skyrmion [26,27], topological superconductivity [28], and intrinsic spin

Hall effect [29]. Thus, the LAO/STO interface offers us a unique playground to explore the Rashba SOC effects and their potential device applications.

Different from bulk STO, where the conduction band edge is contributed by Ti $t_{2g}$ ($d_{xy}$, $d_{xz}$, and $d_{yz}$) orbitals that are degenerate at the Γ point, the conduction band edge of LAO/STO (001) interface is mainly derived from Ti $d_{xy}$ band. This lifted degeneracy with respect to $d_{yz}/d_{xz}$ bands is due to the broken inversion symmetry and confinement effect [3,30,31]. Consequently, it was found that the $d_{xy}$ band and $d_{yz}/d_{xz}$ bands of the LAO/STO interface have different effect on the Rashba SOC. The Rashba SOC near the bottom of $d_{xy}$ band takes a phenomenological linear-in-$k$ form: $H_R = \alpha_R(k \times \sigma) \cdot z$, where the coefficient $\alpha_R$ indicates the Rashba SOC strength, $\sigma$ denotes a vector of Pauli spin matrices, $z$ is a unit vector normal to the interface [3,30–33]. In contrast, for the $d_{yz}/d_{xz}$ bands, the Rashba SOC is cubically dependent on $k$, which even shows anisotropic behaviour [34]. In theoretical framework, SOC and interorbital hopping between Ti $t_{2g}$ orbitals are widely used to explain the Rashba SOC at the LAO/STO interface [3,30,31,35,36]. The crossing of multi $t_{2g}$ bands induces large spin-orbit coupling and interorbital hopping, which is called multiorbital effect. It has been reported that the multiorbital effect contribute to large Rashba SOC at the band crossing region [3,30,31,35,36]. Experimentally, it was found that the value of $\alpha_R$ at the LAO/STO interface is related to carrier density, which is tunable by external electric field [18,36–38]. With the increase of carrier density, $\alpha_R$ at the LAO/STO interface first increases as the Fermi level gradually approaches the $t_{2g}$ band crossing region and then decreases when the Fermi level is above the crossing region. Besides, it was reported that the orbital character at the conduction band edge of the LAO/STO (001) interface can be tuned from $d_{xy}$ to $d_{yz}/d_{xz}$ by an in-plane (parallel to the interface) biaxial compressive strain [39], which in turn changes the relationship between Rashba SOC and $k$.

The electronic properties of the LAO/STO interface are highly dependent on the LAO thickness. Although the Rashba SOC assisted spin to charge currents conversion at the LAO/STO interface has been demonstrated at different LAO thicknesses (2 uc – 40 uc) [19–22], the evolution of Rashba SOC with the LAO thickness is still not well understood. Besides the Rashba SOC, an emerging ferromagnetism has been reported at the conducting LAO/STO interface [40,41], which also induces spin polarization. However, the interplays between interfacial ferromagnetism and Rashba SOC at the LAO/STO interface still await to be unravelled. Thus, in this work, we performed comprehensive first-principles calculations to investigate how the LAO thickness and interface magnetism affect the evolution of Rashba SOC at the LAO/STO (001) interface. We find that the Rashba SOC strength near the conduction band edge decreases as the interface undergoes the insulator-to-metal transition with the increase of LAO thickness. This tendency is ascribed to the LAO-thickness-dependent multiorbital effect, the influence of which near the conduction band edge decreases with the increase of LAO thickness. Moreover, we show that the Rashba SOC is suppressed by the emerging magnetism at the conducting LAO/STO interface.

**Computational details**

All calculations were performed using density-functional theory based Vienna Ab initio Simulation Package (VASP 5.4.4.18) with the Perdew-Burke-Ernzerhof (PBE) functional for the electron exchange-correlation interaction [42–44] and projected augmented wave (PAW) potentials for the interaction between valence electrons and core ions [45]. The on-site coulomb interactions of Ti $d$ orbitals and La $f$ orbitals were taken into account by using the PBE+$U$ method [46]. $U = 5.0$ eV and $J_H = 0.64$ eV were used for Ti 3$d$ orbitals [47,48]. A Hubbard $U$ of 11.0 eV was used for localized $f$ orbitals of La to push them to higher energy position [49]. The plane-wave basis with a cutoff energy of 500 eV was used to expand the electronic wave functions. Γ-centered *k*-point grids for sampling the first Brillouin-zone were set to 12 × 12 × 12 and 12 × 12 × 1 for bulk STO (LAO) and slab models of LAO/STO heterostructures, respectively. A series of slab models (LAO)$_m$/(STO)$_6$ with LaO/TiO$_2$ interfaces [see Fig. 1(a)] were used in calculations to study the effect of LAO thickness, where the thickness of STO substrate is fixed to 6 uc to simulate the STO substrate and the thickness of LAO layers is denoted by *m* unit cells (uc) [10,50]. The value of *m* takes 2, 3, 4, 6, and 8 to capture the insulator-metal transition and thickness effect of the LAO/STO interface. For each slab model, the bottom of the STO substrate was terminated with TiO$_2$ sublayer to avoid the artificial states induced by the bottom SrO layers [51]. To minimize the interaction between neighbouring surfaces, a vacuum layer with a thickness of larger than 20 Å was applied along the *z* direction (out-of-plane). All atoms except those of the bottom TiO$_2$ sublayer were fully relaxed until the force acting on each atom is less than 0.02 eV/Å. The in-plane lattice parameter of the heterostructure was fixed to that of the optimized STO bulk (3.970 Å), which is consistent with previous calculations and the experimentally reported 3.905Å [9,10,16]. To avoid the spurious electric field, dipole corrections were applied in calculations [52].

**Results and discussions**

With the increase of LAO thickness, the LAO/STO interface undergoes an insulator-metal transition at the critical thickness of 4 uc, accompanied by the emergence of interfacial ferromagnetism [see Figs. S1 and S2]. The coexistence of the ferromagnetism and Rashba SOC at the conducting LAO/STO interface makes it complicated to analyse the evolution of Rashba SOC with the LAO thickness, as both the Rashba SOC and the interfacial ferromagnetism induce spin polarization and lift spin degeneracy. To gain a clear understanding of their roles and interplay, we focus first on the evolution of the Rashba SOC with the LAO thickness by leaving out the interfacial ferromagnetism in calculations. After that, we include both the Rashba SOC and the interfacial ferromagnetism, and discuss the interplay between the Rashba SOC and the interfacial ferromagnetism at the conducing LAO/STO interface.

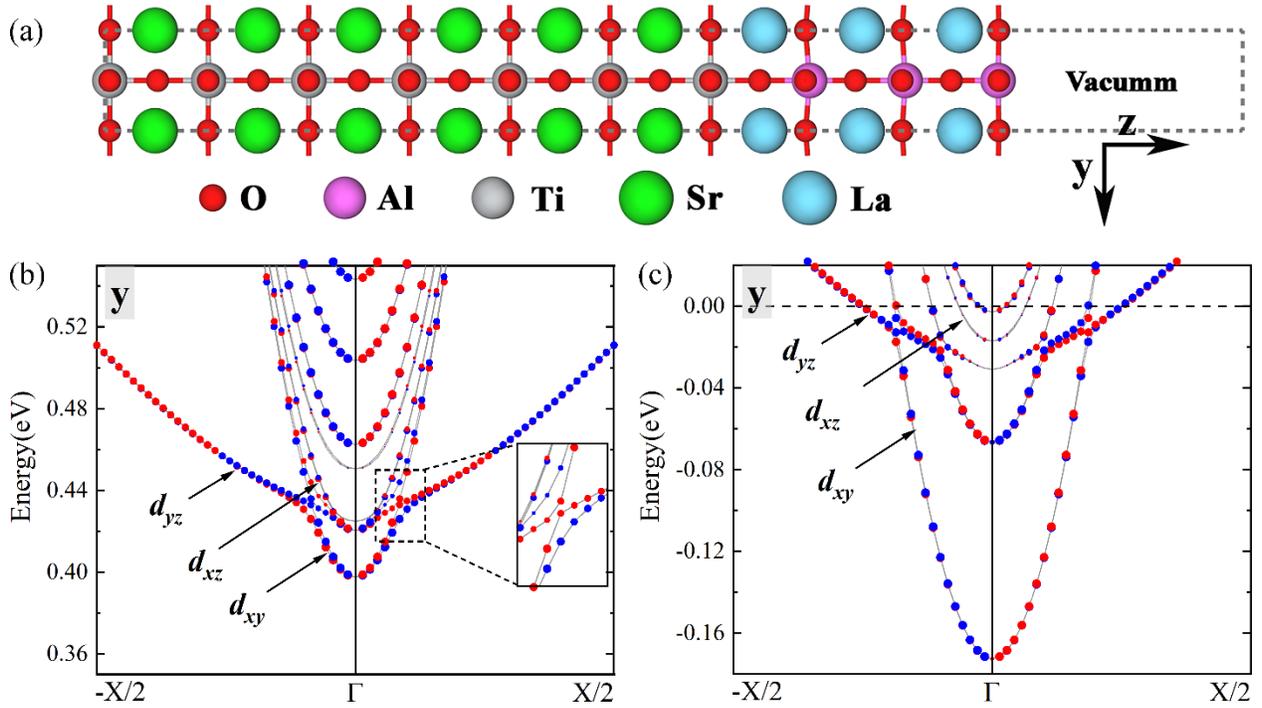

FIG. 1. (a) A schematic diagram of the slab model for the $(LAO)_3/(STO)_6$ heterostructure. SOC included band structures of (b) the $(LAO)_3/(STO)_6$ and (c) $(LAO)_6/(STO)_6$ heterostructures with the projection of electron spins along the *y*-direction. Red and blue dots denote positive and negative components of spin projections, respectively. The inset in (b) is a zoom-in view of the crossing region of the lowest $d_{xy}$ and $d_{yz}/d_{xz}$ bands.

1. **Evolution of Rashba SOC with the increase of LAO thickness**

Figs. 1(b) and 1(c) show SOC-included band structures along $\Gamma - X$ ($\frac{\pi}{a}, 0, 0$) (*a* is the in-plane lattice parameter) of the (LAO)$_3$/(STO)$_6$ and (LAO)$_6$/(STO)$_6$ (001) heterostructures, which are representatives of insulating and conducting LAO/STO interfaces, respectively. For the two band structures, both conduction band edges are contributed by the $d_{xy}$ orbital. A crossing between the $d_{xy}$ and $d_{yz}/d_{xz}$ bands is noticeable near the conduction band edge, as shown in the zoom-in view in the inset of Fig. 1(b). The Rashba SOC induced spin polarization is analysed by projecting spins of electrons along *x*-, *y*-, and *z*-directions. It is found that for both insulating and conducting LAO/STO interfaces, the spins are dominantly polarized along the *y*-direction, which is perpendicular to the selected ***k*** path along the $\Gamma-X$ direction. In contrast, the components of spins along the *x*- and *z*-directions are zero [see Fig. S3]. Fig. 2(a) shows a ***k***-dependent spin splitting ($\Delta_R$) near the conduction band edge (***k*** path from −X/25 to X/25), which is a zoom-in view of the lowest $d_{xy}$ band in Fig. 1(b). Both the ***k***-dependent spin polarization and spin splitting are features of Rashba SOC.

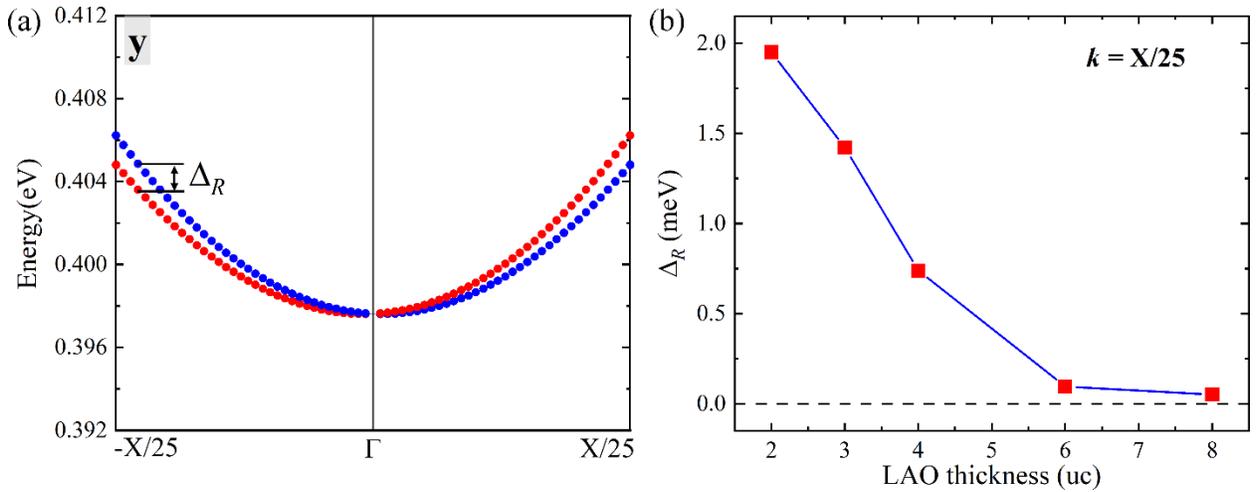

FIG. 2: (a) Rashba spin splitting ($\Delta_R$) near the bottom of the lowest $d_{xy}$ band of the (LAO)$_3$/(STO)$_6$ heterostructure, where red and blue dots denote positive and negative components of spin

projections along the *y*-direction, respectively. The size of the dot indicates the projected weight of the corresponding spin component. (b) The LAO-thickness-dependent $\Delta_R$ of the lowest $d_{xy}$ bands at the X/25.

Fig. 2 (b) shows the evolution of Rashba spin splitting ($\Delta_R$) near the bottom of the lowest $d_{xy}$ band (at the X/25) with respect to the LAO thickness. It can be seen that $\Delta_R$ at the X/25 decreases with the increase of the LAO thickness from 2 uc to 8 uc, especially at the thicknesses that the interfacial insulator-metal transition occurs. This trend suggests that the Rashba SOC near the conduction band edge of the insulating interface is stronger than that of the conducting interface. Further analysis reveals that the tendency of $\Delta_R$ in Fig. 2(b) is relevant to the multiorbital effect of the multi $t_{2g}$ bands [3,36]. In principle, the multiorbital effect is correlated to both SOC and interorbital hopping. [3,30,31,35]. At the crossing region of multi $t_{2g}$ bands, the SOC is prominent. This is because at this region the SOC strongly mixes the orbitals and couples the spin and orbital degree of freedoms pronouncedly [36]. Besides, the interorbital hopping at the band crossing region is also strong [30]. The multiorbital effect therefore leads to large Rashba SOC at the band crossing region [3,30,31,35,36]. The multiorbital effect induced large enhancement of $\Delta_R$ is demonstrated in Fig. 3(a), where the maximum $\Delta_R$ with amplitudes up to several meV occur at the *k*-points around $\pm$X/12 and $\pm$X/6 for insulating (red solid line, $(LAO)_3/(STO)_6$) and conducting (blue solid line, $(LAO)_6/(STO)_6$) interfaces, respectively. These *k*-points correspond to crossing regions of the lowest $d_{xy}$ and $d_{yz}/d_{xz}$ bands for the two interfaces. The enhanced $\Delta_R$ at the band crossing region is consistent with previous studies [3,36]. In contrast, $\Delta_R$ decreases dramatically for the *k*-points that are not at the band crossing regions due to the vanishing multiorbital effect. Fig. 3(a) also reveals that the maximum $\Delta_R$ at the insulating interface (6.9 meV, $(LAO)_3/(STO)_6$) is larger than that at the conducting interface (4.7 meV, $(LAO)_6/(STO)_6$). This suggests a stronger multiorbital effect at the band crossing region of the insulating interface.

Besides, the position of the band crossing region are highly dependent on the band configuration at the LAO/STO interface. Due to the symmetry broken and confinement effect at the interface, the lowest $d_{xy}$ band splits from the $d_{yz}/d_{xz}$ bands at the Γ point, which is denoted by $δ$ in insets in Fig. 3(b). The $δ$ is correlated to the distance between the band crossing region and the bottom of the the lowest $d_{xy}$ band. With the increase of LAO thickness, $δ$ becomes larger [see Fig. 3(b)], as the lowest $d_{xy}$ band is pushed towards lower energy position by the insulator-metal transition accompanied charge transfer [51]. In this sense, the band crossing region moves away from the bottom of the lowest $d_{xy}$ band with the insulator-to-metal transition of the interface. Therefore, the multiorbital effect near the bottom of the lowest $d_{xy}$ band is stronger at the insulating LAO/STO interface than that at the conducting LAO/STO interface. This explains why $Δ_R$ at the X/25 decreases with the increase of the LAO thickness as observed in Fig. 2(b).

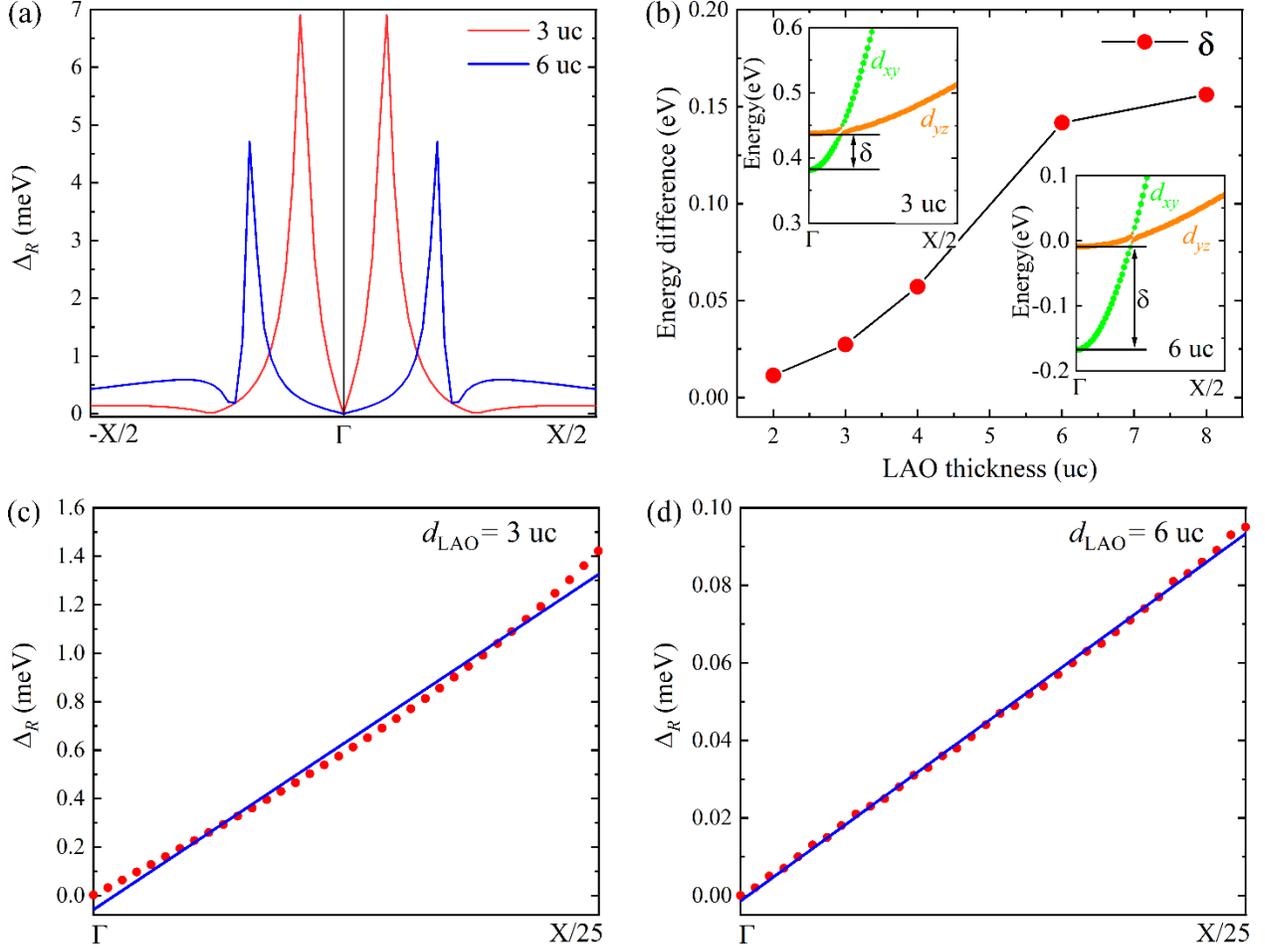

FIG. 3: (a) Rashba spin splitting ($\Delta_R$) of the lowest band along the *k*-path from -X/2 to X/2 for the $(LAO)_3/(STO)_6$ (red solid line) and $((LAO)_6/(STO)_6)$ (blue solid line) heterostructure. (b) Variation of the interface-induced orbital splitting $\delta$ between the lowest $d_{xy}$ and $d_{yz}/d_{xz}$ bands at the $\Gamma$ point for LAO/STO heterostructures with different LAO thicknesses. $\Delta_R$ from the $\Gamma$ point to the X/25 of the lowest $d_{xy}$ band for (c) $(LAO)_3/(STO)_6$ and (d) $(LAO)_6/(STO)_6$ heterostructures are shown by red dotted lines, respectively. The blue lines in (c) and (d) denote the linear fitting of the $\Delta_R$.

In addition to the Rashba spin splitting, the LAO-thickness-dependent multiorbital effect also influences the relationship between Rashba SOC and **k** near the bottom of the lowest $d_{xy}$ band. To examine this effect, we have plotted the $\boldsymbol{k}$ dependent $\Delta_R$ near the bottom of the lowest $d_{xy}$ band for both the insulating and conducting interface in Figs. 3(c) and 3(d). For the insulating LAO/STO interface, since its band crossing region is close to the bottom of the lowest $d_{xy}$ band (smaller $\delta$), the dependence of the $\Delta_R$ near the bottom of the lowest $d_{xy}$ band on $\boldsymbol{k}$ is found nonlinear [see Fig. 3(c)], resulting from the stronger multiorbital effect. For the conducting interface where its band crossing region is away from the bottom of the lowest $d_{xy}$ band (larger $\delta$), $\Delta_R$ is expected to be small and can be fitted linearly dependent on $\boldsymbol{k}$, as shown in Fig. 3(d). In that case, the Rashba coefficient $\alpha_R$ along the $x$-direction, therefore, can be obtained by fitting the $\Delta_R$ to the phenomenological linear-in-$\boldsymbol{k}$ Rashba SOC [3]:

$$\Delta_R = 2\alpha_R k_x \qquad (1)$$

The $\alpha_R$ obtained from the linear fitting is about 1.5 meV·Å for the (LAO)$_6$/(STO)$_6$ heterostructure, which is in line with previous reports that $\alpha_R$ is in the range from several to tens of meV·Å for STO based heterostructures [3,17,31,53].

## 2. The interplay between Rashba SOC and interface magnetic ordering

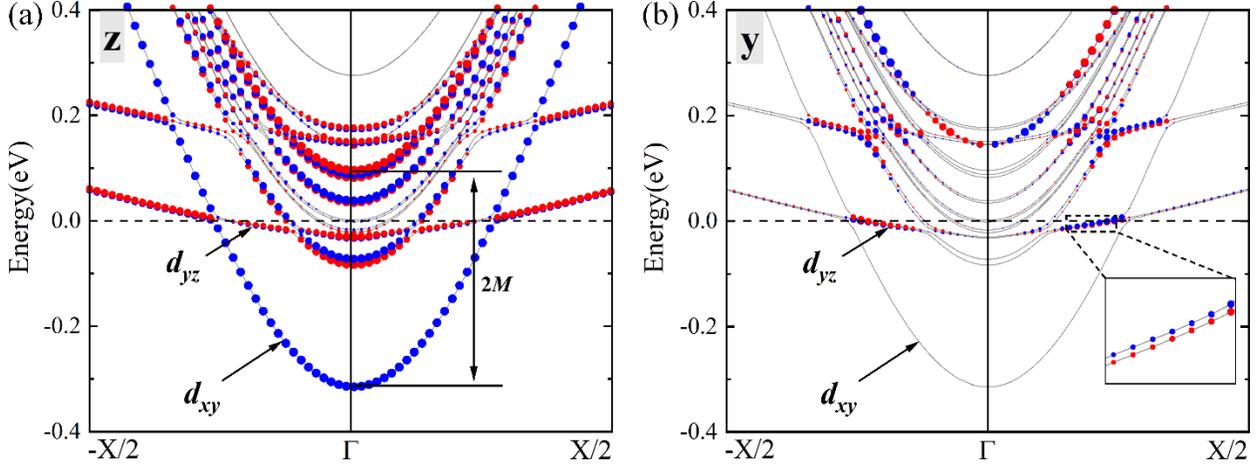

FIG. 4: Band structures of the $(LAO)_6/(STO)_6$ heterostructure including both SOC and interface magnetism. (a) and (b) show band structures with the projection of electron spins along $z$-, and $y$- directions, respectively. Red and blue solid dots denote positive and negative components, respectively. The size of the dot indicates the projected weight of the corresponding spin component.

The above analysis does not include the effect of the interface magnetism. However, along with insulator-metal transition above the critical thickness (LAO thickness $\geq$ 4 uc), the LAO/STO interface energetically favours ferromagnetic ordering. This interfacial magnetic ordering is mainly contributed by the lowest $d_{xy}$ band [see Fig. S2] [15,40,49,54], which induces a large exchange splitting as shown in Fig. 4(a). The interplay between the Rashba SOC and ferromagnetism at the conducting interface is discussed below. The Hamiltonian of the interface including both the interfacial ferromagnetism and the Rashba SOC takes the form [55–57]:

$$H = \frac{\hbar^2 \mathbf{k}^2}{2m^*} + \alpha_R (\mathbf{k} \times \boldsymbol{\sigma}) \cdot \mathbf{z} - M\sigma_z \qquad (2)$$

where $M$ is half of the ferromagnetism induced Zeeman splitting energy, and $\sigma_z$ denotes the interfacial ferromagnetism induced spin component along the $z$-direction (out-of-plane). Here, we assume that the magnetization is along the $z$-direction for simplicity. The coexistence of the ferromagnetism and Rashba SOC leads to two spin-splitting bands:

$$E_\pm = \frac{\hbar^2 \boldsymbol{k}^2}{2m^*} \pm \sqrt{\alpha_R^2 \boldsymbol{k}^2 + M^2} \tag{3}$$

While the ferromagnetism induced spin polarization is along the $z$-direction, the Rashba SOC induced spin polarization is in the $x$-$y$ plane (in-plane) and perpendicular to $\boldsymbol{k}$. The ferromagnetism and the Rashba SOC, therefore, compete on the alignment of spin orientation. The ratio of the spin polarization induced by them can be quantified by:

$$\frac{|\sigma_{//}|}{|\sigma_z|} = \frac{|\boldsymbol{B}_R|}{|\boldsymbol{B}_{FM}|} \tag{4}$$

where $\sigma_{//}$ denotes the Rashba SOC induced spin component in the $x$-$y$ plane. $\boldsymbol{B}_R$ and $\boldsymbol{B}_{FM}$ correspond to Rashba SOC induced effective magnetic field and magnetic field from ferromagnetism, respectively. Figs. 4(a) and 4(b) show that the spins of the bands with large exchange splitting are mainly polarized along the $z$-direction, such as that for the lowest $d_{xy}$ band. In contrast, the Rashba SOC induced spin component (along the $y$-direction) is relatively large for bands with smaller exchange splitting, i.e., the lowest $d_{yz}$ band as shown in the inset of Fig. 4 (b). Experimentally, the magnetism at the LAO/STO interface is complicated and strongly dependent on the sample preparing conditions [11,49,54,58,59]. Both the in-plane and out-of-plane ferromagnetism have been reported at the LAO/STO interface [12,13,60–62]. While the ferromagnetism induced spin polarization is along specific easy axes, the Rashba SOC induced spin polarization is perpendicular to the $\boldsymbol{k}$ of moving electrons. Therefore, Rashba SOC incduced spin orientation is still competed with that induced by the intrinsic ferromagnetism when the magnetic easy axes are in-plane.

In practical applications, the efficiency of the spin and charge currents interconversion is proportional to the strength of the Rashba SOC [19,21,22]. Based on our calculations, the Rashba SOC near the conduction band edge is inversely proportional to the LAO thickness. Below the critical thickness (~4 uc) of the insulator-metal transition, the Rashba SOC near the conduction band edge in the insulating LAO/STO heterostructures is relatively strong. More importantly, the band crossing region with strong Rashba SOC splitting is closer to the conduction band edge, which is highly desired as a much lower doping concentration is needed to access this band crossing region. In contrast, in the conducting LAO/STO heterostructures, the Rashba SOC is relatively weak and barely influenced by the multiorbital effect, as the charge transfer pushes the lowest $d_{xy}$ band towards lower energy [51]. Besides, the Rashba SOC induced spin polarization is suppressed by the charge transfer induced interfacial ferromagnetic ordering. Thus, a LAO/STO heterostructure with thinner LAO is more desirable for the real spintronic applications, such as the spin to charge current conversion with the spin pumping technique [19].

**Conclusions**

In conclusion, by using first-principles calculations, we show that the Rashba spin splitting near the conduction band edge of the LAO/STO interface is tunable with the LAO thickness and reveal an interplay between the Rashba SOC and the interface magnetic ordering. With the increase of LAO thickness, the multiorbital effect induced enhancement to the Rashba SOC near the conduction band edge decreases as the crossing region of the lowest $d_{xy}$ and $d_{yz}/d_{xz}$ bands is gradually away from the bottom of the lowest $d_{xy}$ band. An interplay between Rashba SOC and ferromagnetism occurs at the conducting LAO/STO interface. The spin polarization is dominated by ferromagnetism instead of the Rashba SOC for bands with large exchange splitting. Our results are useful to understand Rashba SOC induced splitting at the complex oxide interface and suggest that the LAO/STO heterostructure with thin LAO thickness is more promising for the spintronic applications.

**References**

[1] A. Manchon, H. C. Koo, J. Nitta, S. M. Frolov, and R. A. Duine, *New Perspectives for Rashba Spin–Orbit Coupling*, Nat. Mater. **14**, 871 (2015).

[2] E. I. Rashba, *Properties of Semiconductors with an Extremum Loop I. Cyclotron and Combinational Resonance in a Magnetic Field Perpendicular to the Plane of the Loop*, Sov. Physics, Solid State **2**, 1109 (1960).

[3] Z. Zhong, A. Tóth, and K. Held, *Theory of Spin-Orbit Coupling at LaAlO3/SrTiO3 Interfaces and SrTiO3 Surfaces*, Phys. Rev. B **87**, 161102 (2013).

[4] F. T. Vas'ko, *Spin Splitting in the Spectrum of Two-Dimensional Electrons Due to the Surface Potential*, Sov. J. Exp. Theor. Phys. Lett. **30**, 541 (1979).

[5] Y. P. Feng, L. Shen, M. Yang, A. Wang, M. Zeng, Q. Wu, S. Chintalapati, and C.-R. Chang, *Prospects of Spintronics Based on 2D Materials*, Wiley Interdiscip. Rev. Comput. Mol. Sci. **7**, e1313 (2017).

[6] J. Varignon, L. Vila, A. Barthélémy, and M. Bibes, *A New Spin for Oxide Interfaces*, Nat. Phys. **14**, 322 (2018).

[7] S. Thiel, *Tunable Quasi-Two-Dimensional Electron Gases in Oxide Heterostructures*, Science (80-. ). **313**, 1942 (2006).

[8] T. C. Asmara, A. Annadi, I. Santoso, P. K. Gogoi, A. Kotlov, H. M. Omer, M. Motapothula, M. B. H. Breese, M. Rübhausen, T. Venkatesan, Ariando, and A. Rusydi, *Mechanisms of Charge Transfer and Redistribution in LaAlO3/SrTiO3 Revealed by High-Energy Optical Conductivity*, Nat. Commun. **5**, 3663 (2014).

[9] J. Zhou, T. C. Asmara, M. Yang, G. A. Sawatzky, Y. P. Feng, and A. Rusydi, *Interplay of

*Electronic Reconstructions, Surface Oxygen Vacancies, and Lattice Distortions in Insulator-Metal Transition of LaAlO3/SrTiO3*, Phys. Rev. B - Condens. Matter Mater. Phys. **92**, 125423 (2015).

[10] A. Ohtomo and H. Y. Hwang, *A High-Mobility Electron Gas at the LaAlO3/SrTiO3 Heterointerface*, Nature **427**, 423 (2004).

[11] Ariando, X. Wang, G. Baskaran, Z. Q. Liu, J. Huijben, J. B. Yi, A. Annadi, A. R. Barman, A. Rusydi, S. Dhar, Y. P. Feng, J. Ding, H. Hilgenkamp, and T. Venkatesan, *Electronic Phase Separation at the LaAlO3/SrTiO3 Interface*, Nat. Commun. **2**, 188 (2011).

[12] J. A. Bert, B. Kalisky, C. Bell, M. Kim, Y. Hikita, H. Y. Hwang, and K. A. Moler, *Direct Imaging of the Coexistence of Ferromagnetism and Superconductivity at the LaAlO3/SrTiO3 Interface*, Nat. Phys. **7**, 767 (2011).

[13] L. Li, C. Richter, J. Mannhart, and R. C. Ashoori, *Coexistence of Magnetic Order and Two-Dimensional Superconductivity at LaAlO3/SrTiO3 Interfaces*, Nat. Phys. **7**, 762 (2011).

[14] Y. Y. Pai, A. Tylan-Tyler, P. Irvin, and J. Levy, *Physics of SrTiO3-Based Heterostructures and Nanostructures: A Review*, Reports Prog. Phys. **81**, 036503 (2018).

[15] A. Brinkman, M. Huijben, M. van Zalk, J. Huijben, U. Zeitler, J. C. Maan, W. G. van der Wiel, G. Rijnders, D. H. A. Blank, and H. Hilgenkamp, *Magnetic Effects at the Interface between Non-Magnetic Oxides*, Nat. Mater. **6**, 493 (2007).

[16] W. Kong, J. Zhou, Y. Z. Luo, T. Yang, S. Wang, J. Chen, A. Rusydi, Y. P. Feng, and M. Yang, *Formation of Two-Dimensional Small Polarons at the Conducting LaAlO$_3$/SrTiO$_3$ Interface*, Phys. Rev. B **100**, 085413 (2019).

[17] A. D. Caviglia, M. Gabay, S. Gariglio, N. Reyren, C. Cancellieri, and J.-M. Triscone,

*Tunable Rashba Spin-Orbit Interaction at Oxide Interfaces*, Phys. Rev. Lett. **104**, 126803 (2010).

[18] G. Herranz, G. Singh, N. Bergeal, A. Jouan, J. Lesueur, J. Gázquez, M. Varela, M. Scigaj, N. Dix, F. Sánchez, and J. Fontcuberta, *Engineering Two-Dimensional Superconductivity and Rashba Spin–Orbit Coupling in LaAlO3/SrTiO3 Quantum Wells by Selective Orbital Occupancy*, Nat. Commun. **6**, 6028 (2015).

[19] E. Lesne, Y. Fu, S. Oyarzun, J. C. Rojas-Sánchez, D. C. Vaz, H. Naganuma, G. Sicoli, J. Attané, M. Jamet, E. Jacquet, J. George, A. Barthélémy, H. Jaffrès, A. Fert, M. Bibes, and L. Vila, *Highly Efficient and Tunable Spin-to-Charge Conversion through Rashba Coupling at Oxide Interfaces*, Nat. Mater. **15**, 1261 (2016).

[20] Y. Pai, A. Tylan-Tyler, P. Irvin, and J. Levy, *Physics of SrTiO 3 -Based Heterostructures and Nanostructures: A Review*, Reports Prog. Phys. **81**, 036503 (2018).

[21] Q. Song, H. Zhang, T. Su, W. Yuan, Y. Chen, W. Xing, J. Shi, J. Sun, and W. Han, *Observation of Inverse Edelstein Effect in Rashba-Split 2DEG between SrTiO 3 and LaAlO 3 at Room Temperature*, Sci. Adv. **3**, e1602312 (2017).

[22] Y. Wang, R. Ramaswamy, M. Motapothula, K. Narayanapillai, D. Zhu, J. Yu, T. Venkatesan, and H. Yang, *Room-Temperature Giant Charge-to-Spin Conversion at the SrTiO 3 –LaAlO 3 Oxide Interface*, Nano Lett. **17**, 7659 (2017).

[23] J. C. R. Sánchez, L. Vila, G. Desfonds, S. Gambarelli, J. P. Attané, J. M. De Teresa, C. Magén, and A. Fert, *Spin-to-Charge Conversion Using Rashba Coupling at the Interface between Non-Magnetic Materials*, Nat. Commun. **4**, 2944 (2013).

[24] K. Shen, G. Vignale, and R. Raimondi, *Microscopic Theory of the Inverse Edelstein Effect*, Phys. Rev. Lett. **112**, 096601 (2014).

[25] P. Deorani, J. Son, K. Banerjee, N. Koirala, M. Brahlek, S. Oh, and H. Yang, *Observation of Inverse Spin Hall Effect in Bismuth Selenide*, Phys. Rev. B **90**, 94403 (2014).

[26] S. Banerjee, O. Erten, and M. Randeria, *Ferromagnetic Exchange, Spin–Orbit Coupling and Spiral Magnetism at the LaAlO3/SrTiO3 Interface*, Nat. Phys. **9**, 626 (2013).

[27] X. Li, W. V. Liu, and L. Balents, *Spirals and Skyrmions in Two Dimensional Oxide Heterostructures*, (2014).

[28] S. Nakosai, Y. Tanaka, and N. Nagaosa, *Topological Superconductivity in Bilayer Rashba System*, (2012).

[29] L. X. Hayden, R. Raimondi, M. E. Flatté, and G. Vignale, *Intrinsic Spin Hall Effect at Asymmetric Oxide Interfaces: Role of Transverse Wave Functions*, Phys. Rev. B **88**, 75405 (2013).

[30] Y. Kim, R. M. Lutchyn, and C. Nayak, *Origin and Transport Signatures of Spin-Orbit Interactions in One-and Two-Dimensional SrTiO 3 -Based Heterostructures*, Phys. Rev. B **8720**, (2013).

[31] G. Khalsa, B. Lee, and A. H. MacDonald, *Theory of T2g Electron-Gas Rashba Interactions*, Phys. Rev. B **88**, 041302 (2013).

[32] J. Zhou, W.-Y. Shan, and D. Xiao, *Spin Responses and Effective Hamiltonian for the Two-Dimensional Electron Gas at the Oxide Interface LaAlO 3 /SrTiO 3*, RAPID Commun. Phys. Rev. B **91**, 241302 (2015).

[33] C. S. Ho, W. Kong, M. Yang, A. Rusydi, and M. B. A. Jalil, *Tunable Spin and Orbital Polarization in SrTiO 3 -Based Heterostructures*, New J. Phys. **21**, 103016 (2019).

[34] H. J. Harsan Ma, J. Zhou, M. Yang, Y. Liu, S. W. Zeng, W. X. Zhou, L. C. Zhang, T. Venkatesan, and Y. P. Feng, *Giant Crystalline Anisotropic Magnetoresistance in*

*Nonmagnetic Perovskite Oxide Heterostructures*, Phys. Rev. B **95**, 155314 (2017).

[35] P. D. C. King, S. M. K. Walker, A. Tamai, A. De La Torre, T. Eknapakul, P. Buaphet, S. K. Mo, W. Meevasana, M. S. Bahramy, and F. Baumberger, *Quasiparticle Dynamics and Spin-Orbital Texture of the SrTiO3 Two-Dimensional Electron Gas*, Nat. Commun. **5**, 1 (2014).

[36] A. Joshua, J. Ruhman, and E. Altman, *A Universal Critical Density Underlying the Physics of Electrons at the LaAlo 3 /SrTio 3 Interface*, Nat. Commun. **3**, (2012).

[37] A. D. Caviglia, M. Gabay, S. Gariglio, N. Reyren, C. Cancellieri, and J.-M. Triscone, *Tunable Rashba Spin-Orbit Interaction at Oxide Interfaces*, (n.d.).

[38] C. Yin, P. Seiler, L. M. K. Tang, I. Leermakers, N. Lebedev, U. Zeitler, and J. Aarts, *Tuning Rashba Spin-Orbit Coupling at LaAlO$_3$/SrTiO$_3$ Interfaces by Band Filling*, Phys. Rev. B **101**, 245114 (2020).

[39] W. Lin, L. Li, F. Doğan, C. Li, H. Rotella, X. Yu, B. Zhang, Y. Li, W. S. Lew, S. Wang, W. Prellier, S. J. Pennycook, J. Chen, Z. Zhong, A. Manchon, and T. Wu, *Interface-Based Tuning of Rashba Spin-Orbit Interaction in Asymmetric Oxide Heterostructures with 3d Electrons*, Nat. Commun. **10**, 3052 (2019).

[40] J.-S. Lee, Y. W. Xie, H. K. Sato, C. Bell, Y. Hikita, H. Y. Hwang, and C.-C. Kao, *Titanium Dx y Ferromagnetism at the LaAlO3/SrTiO3 Interface*, Nat. Mater. **12**, 703 (2013).

[41] K. Michaeli, A. C. Potter, and P. A. Lee, *Superconducting and Ferromagnetic Phases in SrTiO3/LaAlO3 Oxide Interface Structures: Possibility of Finite Momentum Pairing*, Phys. Rev. Lett. **108**, 117003 (2012).

[42] G. Kresse and J. Hafner, *Ab Initio Molecular Dynamics for Liquid Metals*, Phys. Rev. B

**47**, 558 (1993).

[43] G. Kresse and J. Hafner, *Ab Initio Molecular Dynamics for Open-Shell Transition Metals*, Phys. Rev. B **48**, 13115 (1993).

[44] G. Kresse and J. Furthmüller, *Efficiency of Ab-Initio Total Energy Calculations for Metals and Semiconductors Using a Plane-Wave Basis Set*, Comput. Mater. Sci. **6**, 15 (1996).

[45] G. Kresse and D. Joubert, *From Ultrasoft Pseudopotentials to the Projector Augmented-Wave Method*, Phys. Rev. B **59**, 1758 (1999).

[46] A. I. Liechtenstein, V. I. Anisimov, and J. Zaanen, Density-Functional Theory and Strong Interactions: Orbital Ordering in Mott-Hubbard Insulators, 1995.

[47] M. Altmeyer, H. O. Jeschke, O. Hijano-Cubelos, C. Martins, F. Lechermann, K. Koepernik, A. F. Santander-Syro, M. J. Rozenberg, R. Valentí, and M. Gabay, *Magnetism, Spin Texture, and In-Gap States: Atomic Specialization at the Surface of Oxygen-Deficient $SrTiO_3$*, Phys. Rev. Lett. **116**, 157203 (2016).

[48] S. Okamoto, A. J. Millis, and N. A. Spaldin, *Lattice Relaxation in Oxide Heterostructures: LaTiO 3 =SrTiO 3 Superlattices*, (2006).

[49] M. Yang, Ariando, J. Zhou, T. C. Asmara, P. Krüger, X. J. Yu, X. Wang, C. Sanchez-Hanke, Y. P. Feng, T. Venkatesan, and A. Rusydi, *Direct Observation of Roomerature Stable Magnetism in LaAlO3/SrTiO3 Heterostructures*, ACS Appl. Mater. Interfaces **10**, 9774 (2018).

[50] N. Reyren, S. Thiel, A. D. Caviglia, L. F. Kourkoutis, G. Hammerl, C. Richter, C. W. Schneider, T. Kopp, A.-S. Ruetschi, D. Jaccard, M. Gabay, D. A. Muller, J.-M. Triscone, and J. Mannhart, *Superconducting Interfaces Between Insulating Oxides*, Science (80-. ).

**317**, 1196 (2007).

[51] W.-J. Son, E. Cho, B. Lee, J. Lee, and S. Han, *Density and Spatial Distribution of Charge Carriers in the Intrinsic N-Type LaAlO 3-SrTiO 3 Interface*, (n.d.).

[52] L. Bengtsson, Dipole Correction for Surface Supercell Calculations, n.d.

[53] P. D. C. King, S. McKeown Walker, A. Tamai, A. de la Torre, T. Eknapakul, P. Buaphet, S. Mo, W. Meevasana, M. S. Bahramy, and F. Baumberger, *Quasiparticle Dynamics and Spin–Orbital Texture of the SrTiO3 Two-Dimensional Electron Gas*, Nat. Commun. **5**, 3414 (2014).

[54] L. Yu and A. Zunger, *A Polarity-Induced Defect Mechanism for Conductivity and Magnetism at Polar–Nonpolar Oxide Interfaces*, Nat. Commun. **5**, 5118 (2014).

[55] K. Rahmanizadeh, G. Bihlmayer, and S. Blügel, *Spin-Orbit and Exchange Effects in the 2DEG of BiAlO 3 -Based Oxide Heterostructures*, EPL (Europhysics Lett. **115**, 17006 (2016).

[56] D. Oshima, K. Taguchi, and Y. Tanaka, *Unconventional Gate Voltage Dependence of the Charge Conductance Caused by Spin-Splitting Fermi Surface by Rashba-Type Spin-Orbit Coupling*, (2019).

[57] P. Středa and P. S. ˇEba, *Antisymmetric Spin Filtering in One-Dimensional Electron Systems with Uniform Spin-Orbit Coupling*, (n.d.).

[58] F. Bi, M. Huang, S. Ryu, H. Lee, C.-W. Bark, C.-B. Eom, P. Irvin, and J. Levy, *Room-Temperature Electronically-Controlled Ferromagnetism at the LaAlO3/SrTiO3 Interface*, Nat. Commun. **5**, 5019 (2014).

[59] F. Bi, M. Huang, H. Lee, C.-B. Eom, P. Irvin, and J. Levy, *LaAlO 3 Thickness Window for Electronically Controlled Magnetism at LaAlO 3 /SrTiO 3 Heterointerfaces*, Appl. Phys.

Lett. **107**, 082402 (2015).

[60] D. S. Park, A. D. Rata, I. V. Maznichenko, S. Ostanin, Y. L. Gan, S. Agrestini, G. J. Rees, M. Walker, J. Li, J. Herrero-Martin, G. Singh, Z. Luo, A. Bhatnagar, Y. Z. Chen, V. Tileli, P. Muralt, A. Kalaboukhov, I. Mertig, K. Dörr, A. Ernst, and N. Pryds, *The Emergence of Magnetic Ordering at Complex Oxide Interfaces Tuned by Defects*, Nat. Commun. **11**, 1 (2020).

[61] T. D. N. Ngo, J.-W. Chang, K. Lee, S. Han, J. S. Lee, Y. H. Kim, M.-H. Jung, Y.-J. Doh, M.-S. Choi, J. Song, and J. Kim, *Polarity-Tunable Magnetic Tunnel Junctions Based on Ferromagnetism at Oxide Heterointerfaces*, Nat. Commun. **6**, 8035 (2015).

[62] B. Kalisky, J. A. Bert, C. Bell, Y. Xie, H. K. Sato, M. Hosoda, Y. Hikita, H. Y. Hwang, and K. A. Moler, *Scanning Probe Manipulation of Magnetism at the LaAlO 3 /SrTiO 3 Heterointerface*, (2012).